\documentstyle[11pt,paspconf]{article}

\newcommand{\apg}{\:^{>}_{\sim}\:}
\newcommand{\apl}{\:^{<}_{\sim}\:}

\begin{document}

\title{The Stony Brook Photometric Redshifts of Faint Galaxies in the Hubble
Deep Fields}

\author{Kenneth M. Lanzetta\altaffilmark{1}, Hsiao-Wen Chen\altaffilmark{1},
Alberto Fern\'andez-Soto\altaffilmark{2}, Sebastian Pascarelle\altaffilmark{1},
Rick Puetter\altaffilmark{3}, Noriaki Yahata\altaffilmark{1}, and Amos
Yahil\altaffilmark{1}}

\altaffiltext{1}
{Department of Physics and Astronomy, State University of New York at Stony
Brook, Stony Brook, NY 11794-3800, U.S.A.}

\altaffiltext{2}
{Department of Astrophysics and Optics, School of Physics, University of New
South Wales, Kensington--Sydney, NSW 2052, AUSTRALIA}

\altaffiltext{3}
{Center for Astrophysics and Space Sciences, University of California at San
Diego, La Jolla, CA 92093--0424, U.S.A.}

\begin{abstract}

  We report on some aspects of the current status of our efforts to establish
properties of faint galaxies by applying our photometric redshift technique to
faint galaxies in the HDF and HDF-S WFPC2 and NICMOS fields.

\end{abstract}

\keywords{high-redshift galaxies, star formation history}

\section{Introduction}

  Over the past several years, we have applied our photometric redshift
technique to faint galaxies in the Hubble Deep Field (HDF) and Hubble Deep
Field South (HDF-S) WFPC2 and NICMOS fields.  Our objective is to establish
properties of galaxies that are too faint to be spectroscopically identified by
the largest ground-based telescopes.  Our experiences indicate that photometric
redshift measurements are at least as robust and reliable as spectroscopic
redshift measurements (and probably more so).  The photometric redshift
technique thus provides a means of obtaining redshift identifications of large
samples of faint galaxies.  Here we report on some aspects of the current
status of our efforts.

\section{Observations and Analysis}

  Our current observations and analysis differ from our previous observations
and analysis in three ways:  First, we have included all publicly available
ground- and space-based imaging observations of the HDF, HDF-S WFPC2, and HDF-S
NICMOS fields.  Details of the observations are summarized in Table 1.  Second,
we have developed and applied a new quasi-optimal photometry technique based on
fitting models of the spatial profiles of the objects (which are obtained using
a non-negative least squares image reconstruction method) to the ground- and
space-based images according to the spatial profile fitting technique described
previously by Fern\'andez-Soto, Lanzetta, \& Yahil (1999).  For faint objects,
the signal-to-noise ratios obtained by our new photometry technique are larger
than the signal-to-noise ratios obtained by aperture photometry techniques by
typically a factor of two.  Third, we have measured photometric redshifts using
a sequence of six spectrophotometric templates, including the four templates of
our previous analysis (of E/S0, Sbc, Scd, and Irr galaxies) and two new
templates (of star-forming galaxies).  Inclusion of the two new templates
eliminates the tendency of our previous analysis to systematically
underestimate the redshifts of galaxies of redshift $2 < z < 3$ (by a redshift
offset of roughly 0.3), in agreement with results found previously by Ben\'itez
et al.\ (1999).

\begin{center}
\begin{tabular}{p{1.75in}l}
\multicolumn{2}{c}{Table 1} \\
\hline
\hline
\multicolumn{1}{c}{Field} & \multicolumn{1}{c}{Filters} \\
\hline
HDF \dotfill          & F300W, F450W, F606W, F814W, \\
& F110W, F160W, $J$, $H$, $K$ \\
HDF-S WFPC2 \dotfill  & F300W, F450W, F606W, F814W, \\
& $U$, $B$, $V$, $R$, $I$, $J$, $H$, $K$ \\
HDF-S NICMOS \dotfill & F110W, F160W, F222M, STIS, \\
& $U$, $B$, $V$, $R$, $I$ \\
\hline
\end{tabular}
\end{center}

  The accuracy and reliability of the photometric redshift technique is
illustrated in Figure 1, which shows the comparison of 108 photometric and
reliable spectroscopic redshifts in HDF and HDF-S.  (Note that a non-negligible
fraction of published spectroscopic redshift measurements of galaxies in HDF
and HDF-S have been shown to be in error and so must be excluded from
consideration.)  With the sequence of six spectrophotometric templates, the
photometric redshifts are accurate to within an RMS relative uncertainty of
$\Delta z/(1 + z) \apl 10\%$ at all redshifts $z < 6$ that have as yet been
examined.

\begin{figure}[th]
\includegraphics{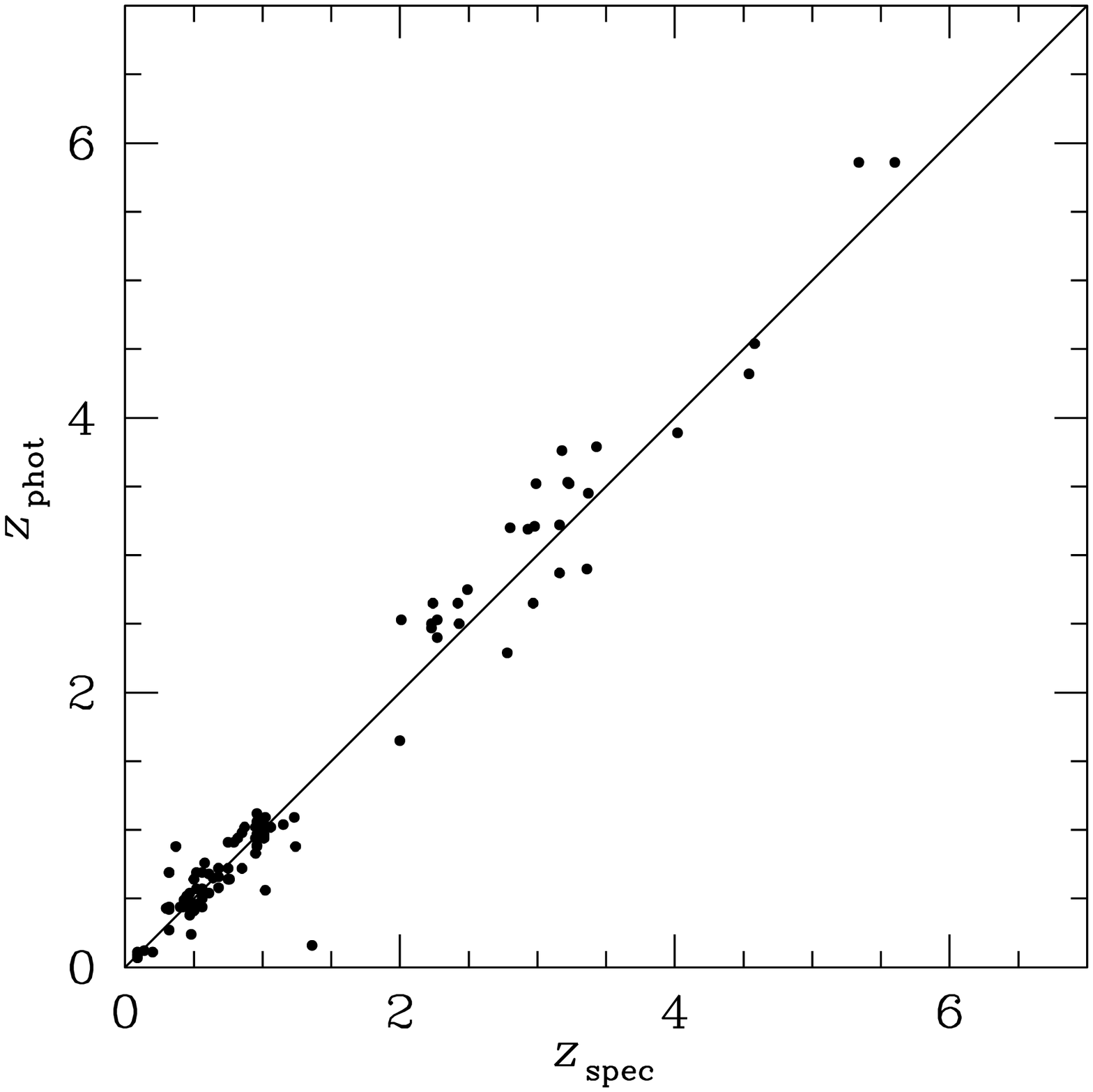}
\vspace{2.65in}
\caption{Comparison of 108 photometric and reliable spectroscopic measurements
of galaxies in HDF and HDF-S.  The RMS dispersion between the photometric and
reliable spectroscopic measurements is $\approx 0.1$ at $z < 2$, $\approx 0.3$
at $2 < z < 4$, and $\approx 0.15$ at $z > 4$.}
\end{figure}

\section{Stony Brook Faint Galaxy Redshift Survey}

  Our analysis of the HDF and HDF-S WFPC2 and NICMOS fields constitutes a
survey of galaxies to the faintest energy flux density and surface brightness
limits currently accessible.  Properties of the survey are as follows:

  First, we have determined nine- or 12-band photometric redshifts of galaxies
in three fields.  Second, we have selected galaxies at both optical and
infrared wavelengths, in two or more of the F814W, F160W, $H$, and $K$ bands
(depending on field).  Third, we have characterized the survey area versus
depth relations, as functions of both energy flux density and surface
brightness.  Fourth, we have established properties of the extremely faint
galaxy population using a maximum-likelihood parameter estimation technique
and a bootstrap resampling parameter uncertainty estimation technique.  The
parameter uncertainties explicitly account for the effects of photometric
error, sampling error, and cosmic dispersion with respect to the
spectrophotometric templates.  The Stony Brook faint galaxy redshift survey
includes nearly 3000 faint galaxies, of which 671 galaxies are of redshift $z >
2$.

\section{Some High (and Not So High) Redshift Galaxies}

  Examples of some high and not so high redshift galaxies are shown in Figure
2, which shows the observed and modeled spectral energy distributions and
redshift likelihood functions of galaxies A through C, which we previously
identified as candidate extremely high redshift galaxies on the basis of 
ground-based near-infrared measurements (Lanzetta, Yahil, \& Fern\'andez-Soto
1998).  Our current analysis indicates that galaxy A is probably an early-type
galaxy of redshift $z = 2.34 \pm 0.36$ and that galaxies B and C are probably
star-forming galaxies of redshift $z > 13$.  (An extremely high redshift
interpretation of galaxy A is apparently ruled out by faint but significant
energy flux density in the F606W and F814W filters.  A low-redshift, highly
obscured and reddened interpretation of galaxies B and C is apparently ruled
out by the large flux decrement between the $K$ and F160W images.)

\begin{figure}[th]
\includegraphics{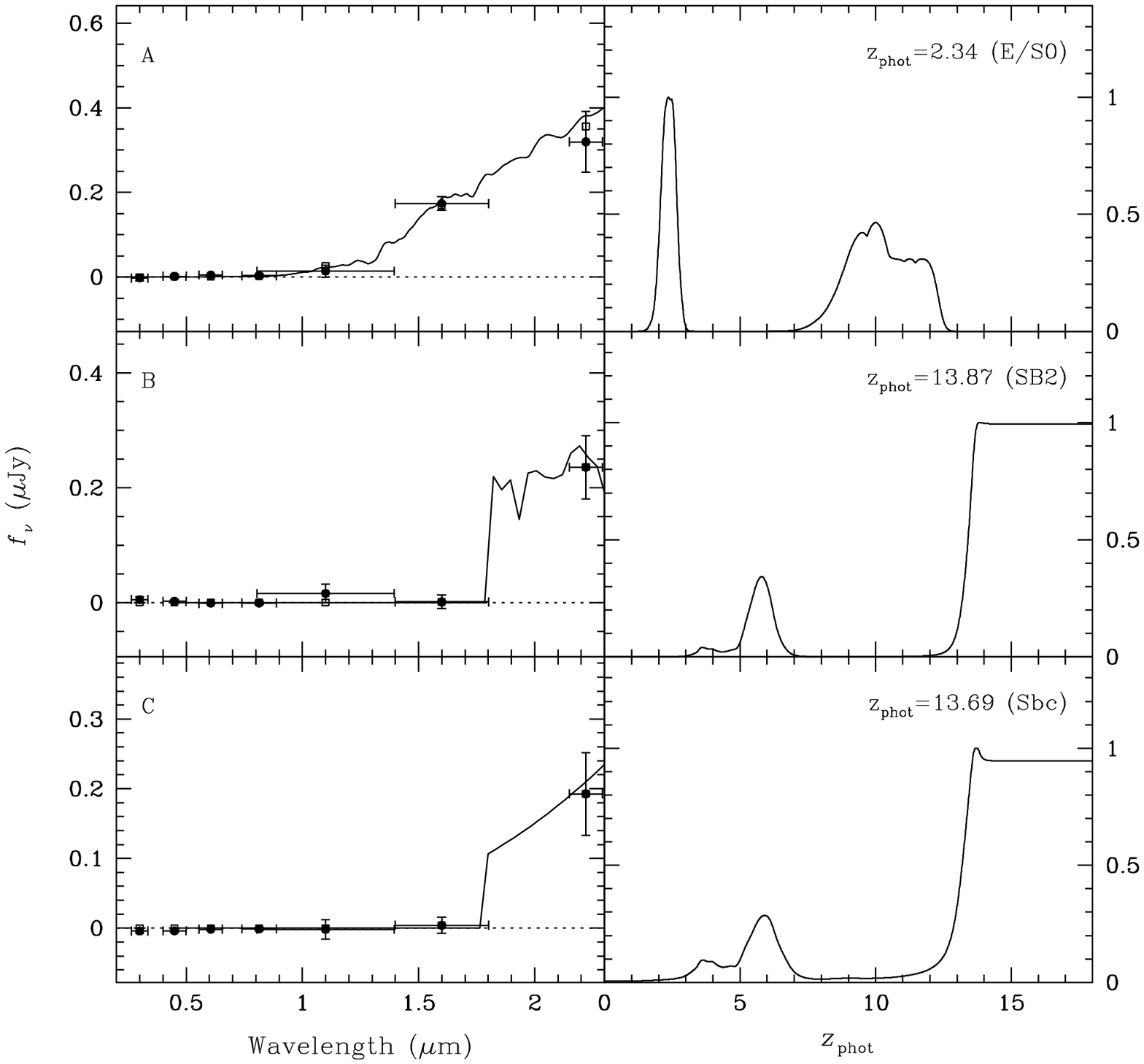}
\vspace{3.55in}
\caption{Observed and modeled spectral energy distributions (left panels) and
redshift likelihood functions (right panels) of galaxies A through C.}
\end{figure}

\section{The Galaxy Luminosity Function at $z > 2$}

  We have modeled the rest-frame 1500 \AA\ luminosity function of galaxies of
redshift $z > 2$ by adopting an evolving Schechter luminosity function
\begin{equation}
\Phi(L,z) = \Phi_* / L_*(z) [ L / L_*(z) ]^{-\alpha} \exp[ -L / L_* (z)]
\end{equation}
with
\begin{equation}
L_*(z) = L_*(z = 3) \left( \frac{1 + z}{4} \right)^\beta.
\end{equation}
The best-fit parameters for a simultaneous fit to the HDF and HDF-S WFPC2 and
NICMOS fields (where we have related selection in different bands by adopting
the spectral energy distribution of a star-forming galaxy) are $\Phi_* = 0.004
\pm 0.001 \ h^3$ Mpc$^{-3}$, $L_* = 2.7 \pm 0.3 \times 10^{28} \ h^{-2}$ erg
s$^{-1}$ Hz$^{-1}$, $\alpha = 1.49 \pm 0.03$, and $\beta = -1.2 \pm 0.3$.  The
best-fit model is compared with the observations in Figure 3, which shows the
cumulative galaxy surface density versus redshift and magnitude for galaxies
selected in the F814W and F160W bands.
    
\begin{figure}[th]
\includegraphics{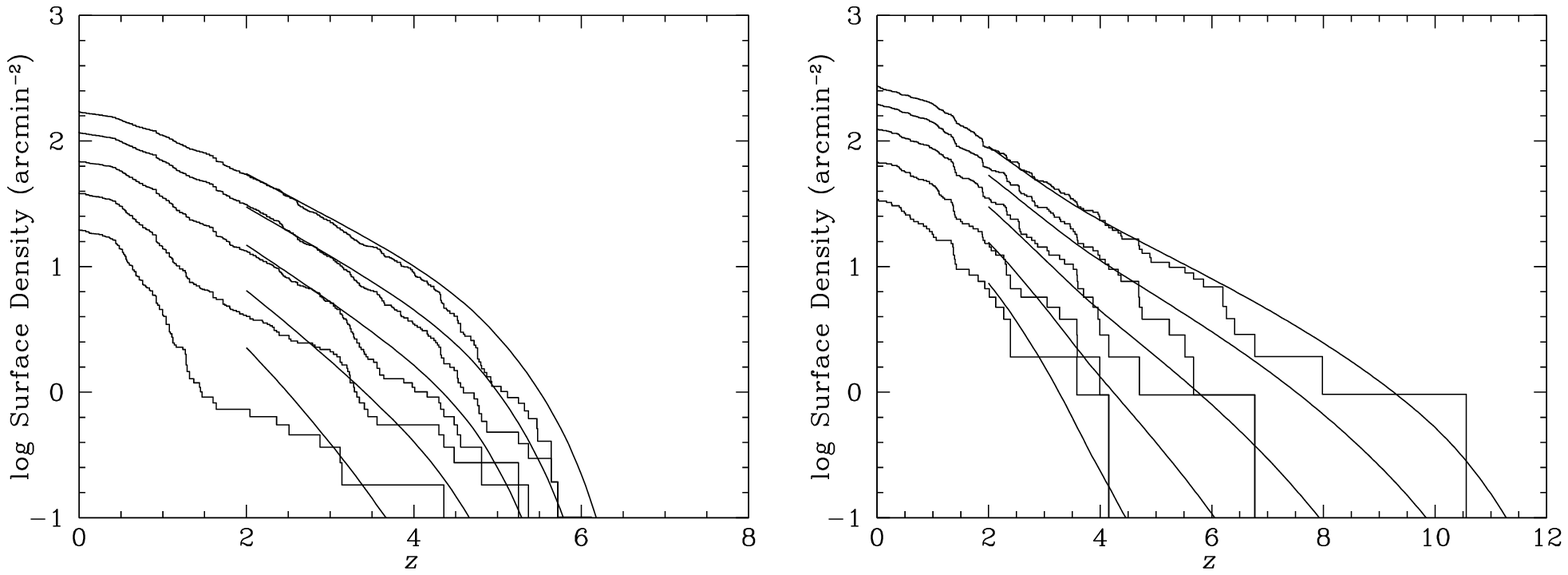}
\vspace{1.65in}
\caption{Cumulative galaxy surface density versus redshift and magnitude (i.e.\
surface density of galaxies of redshift greater than a given redshift) for
galaxies selected in the F814W (left panel) and F160W (right panel) bands.
Smooth curves are best-fit model, and jagged curves are observations.
Different curves show different magnitude thresholds, ranging from $AB = 24$
(bottom curves) through $AB = 28$ (top curves).}
\end{figure}

\section{Effects of Cosmological Surface Brightness Dimming}

  Results of the previous section indicate that the galaxy luminosity function
is only mildly evolving at redshifts $z > 2$, i.e.\ as $(1 + z)^\beta$ with
$\beta \approx -1$.  But due to $(1 + z)^3$ cosmological surface brightness
dimming, the measured luminosity of extended objects will decrease with
increasing redshift, even if the luminosities of the objects remain constant.
For this reason, we consider it almost meaningless to interpret the galaxy
luminosity function (or its moments) over a redshift interval spanning $z = 2$
through $z = 10$, at least without explicitly taking account of surface
brightness effects.

  To make explicit the effects of cosmological surface brightness dimming on
observations of high-redshift galaxies, we have constructed the ``star
formation rate intensity distribution function'' $h(x)$.  Specifically, we
consider all pixels contained within galaxies on an individual pixel-by-pixel
basis.  Given the redshift of a pixel (which is set by the photometric redshift
of the host galaxy), an empirical $k$ correction (which is set by the model
spectral energy distribution of the host galaxy) and a cosmological model
determine the rest-frame 1500 \AA\ luminosity of the pixel, and an angular
plate scale and a cosmological model determine the proper area of the pixel.
Adopting a Salpeter initial mass function to convert the rest-frame 1500 \AA\
luminosity to the star formation rate and dividing the star formation rate by
the proper area yields the ``star formation rate intensity'' $x$ of the pixel.
Summing the proper areas of all pixels within given star formation rate
intensity and redshift intervals, dividing by the star formation rate intensity
interval, and dividing by the comoving volume then yields the ``star formation
rate intensity distribution function,'' which we designate as $h(x)$.  The star
formation rate intensity distribution function $h(x)$ is exactly analogous to
the QSO absorption line systems column density distribution function $f(N)$ (as
a function of neutral hydrogen column density $N$).  In terms of the star
formation rate intensity distribution function, the unobscured cosmic star
formation rate density $\dot{\rho}_s$ (or equivalently the rest-frame
ultraviolet luminosity density) is given by
\begin{equation}
\dot{\rho}_s = \int_0^\infty x h(x) dx.
\end{equation}

  Results are shown in Figure 4, which plots the star formation rate intensity
distribution function $h(x)$ versus star formation rate intensity $x$
determined from galaxies identified in the HDF and HDF-S NICMOS field.  Several
results are apparent on the basis of Figure 4:  First, the star formation rate
intensity threshold of the survey is an extremely strong function of redshift,
ranging from $x_{\rm min} \approx 5 \times 10^{-4}$ $M_\odot$ yr$^{-1}$
kpc$^{-2}$ at $z \approx 0.5$ to $x_{\rm min} \approx 1$ $M_\odot$ yr$^{-1}$
kpc$^{-2}$ at $z \approx 6$.  Second, at redshifts $z \apl 1.5$ [at which
$h(x)$ is measured over a wide range in $x$], the distribution is characterized
by a relatively shallow slope at $\log x \apl -1.5$ $M_\odot$ yr$^{-1}$
kpc$^{-2}$ and by a relatively steep slope at $\log x \apg -1.5$ $M_\odot$
yr$^{-1}$ kpc$^{-2}$.  These slopes are such that the bulk of the cosmic star
formation rate density occurs at $\log x \approx 1.5$ $M_\odot$ yr$^{-1}$
kpc$^{-2}$, which is measured only at redshifts $z \apl 2$.  {\em We conclude
that the cosmic star formation rate density (or equivalently the rest-frame
ultraviolet luminosity density) has not yet been measured at redshifts $z \apg
2$.}  Third, the comoving volume density of the highest star formation rate
intensity regions increases monotonically with increasing redshift.  {\em We
conclude that the comoving volume density of the most intense star formation
regions increases monotonically with increasing redshift} (see also Pascarelle,
Lanzetta, \& Fern\'andez-Soto 1998).

\begin{figure}[th]
\includegraphics{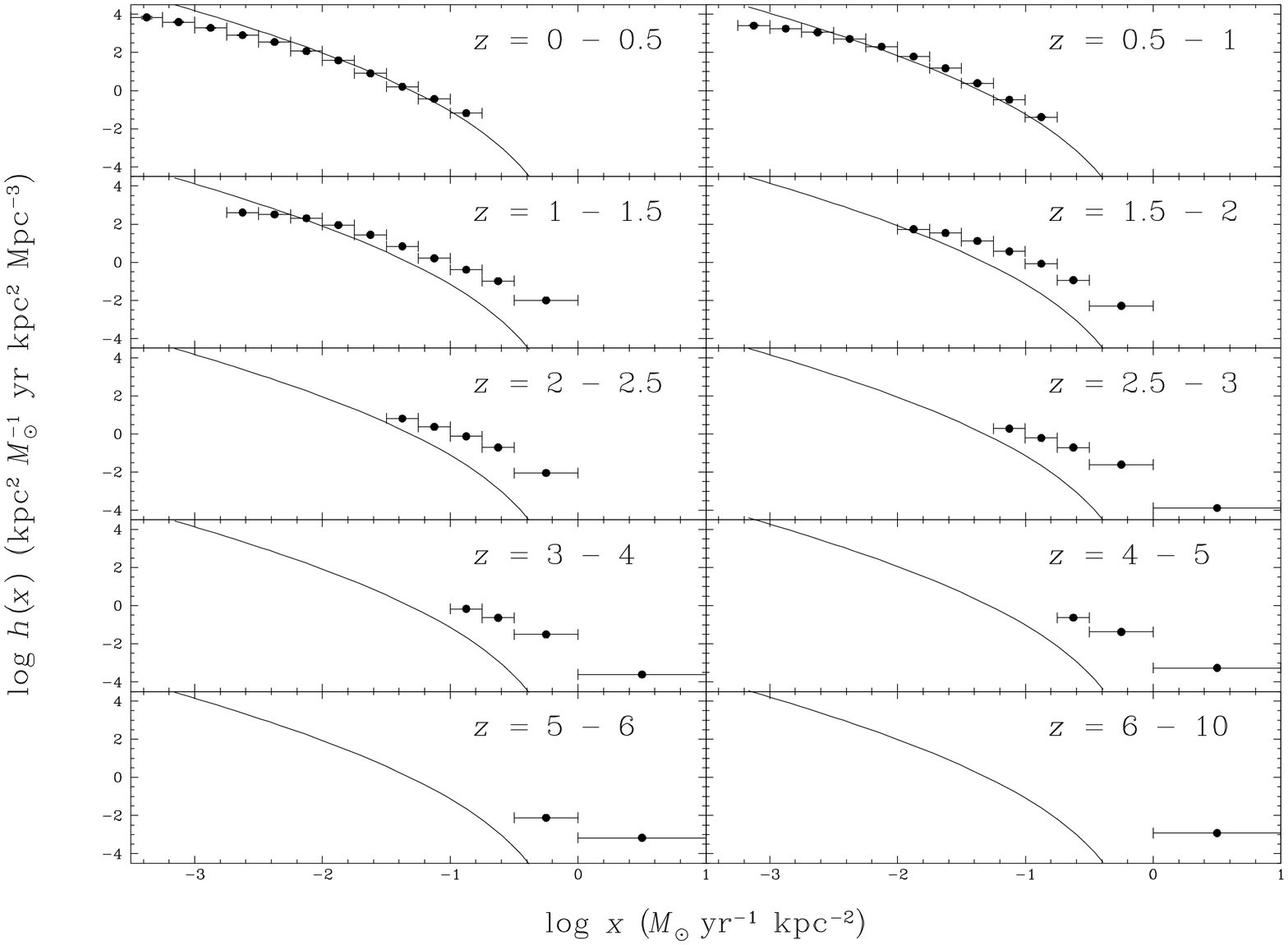}
\vspace{3.40in}
\caption{Logarithm of star formation rate intensity distribution function
$h(x)$ versus logarithm of star formation rate intensity $x$, determined from
galaxies identified in the HDF and HDF-S NICMOS field.  Different panels show
different redshift intervals, ranging from $z = 0$ through 10.  Points show
observations, with vertical error bars indicating $1 \sigma$ uncertainties and
horizontal error bars indicating bin sizes.  Smooth curves show a fiducial
model (based on a bulge spatial profile) adjusted to roughly match the
observations at $z = 0 - 0.5$.}
\end{figure}

\acknowledgments

  We thank Hy Spinrad and Daniel Stern for providing spectroscopic redshift
measurements in advance of publication and acknowledge Mark Dickinson and Roger
Thompson for obtaining NICMOS observations of HDF.  This research was supported
by NASA grant NACW--4422 and NSF grant AST--9624216 and is based on
observations with the NASA/ESA Hubble Space Telescope and on observations
collected at the European Southern Observatory.

\end{document}